# DFT-based energy shifts screening of $Na_xK_{55-x}$ alloy clusters

Jing Zhao, Maolin Bo*

*Chongqing Key Laboratory of Extraordinary Bond Engineering and Advanced Materials Technology (EBEAM), Yangtze Normal University, Chongqing 408100, China*

*E-mail:bmlwd @yznu.edu.cn*

**Abstract**

Compositional effects in $Na_xK_{55-x}$ alloy clusters have been studied using bond-order–length–strength (BOLS) notation and density functional theory (DFT) calculations. The results reveal binding energy shifts of the $Na_xK_{55-x}$ alloy clusters under different elemental compositions. Atomic arrangements that can be used to predict the structures of stable experimental $Na_xK_{55-x}$ alloys were also obtained. Our study of these alloy nanoclusters has uncovered a trend correlating atomic position and composition with binding energy. We believe this data will help in the experimental preparation of alloy nanoclusters.

**Keywords:** NaK alloy cluster, binding energy shifts, density functional theory calculations

# 1. Introduction

Alloy nanoclusters are widely studied not only because of size effects but also because their physical and chemical properties are subject to compositional effects[1]. Size effects influence the proportion of surface atoms with low coordination numbers, which in turn affects magnetic[2, 3], thermal[4, 5], catalytic[6, 7], optical[8] and electronic properties[9]. However, the effects of compositional variations are not limited to the distribution of surface atoms, as is the case for size effects, but they also influence atomic arrangement within cluster nanomaterials[10]. In recent years, in order to understand the performance of nanoalloys at a fundamental level, a large number of experimental[11, 12] and theoretical[13-17] studies have been conducted. In particular, concerning the NaK alloy, Tchaplyguine et al. studied self-assembled clusters using X-ray photoelectron spectroscopy (XPS)[18]. Furthermore, Aguado et al. reported global-minimum structures of 55-atom NaK alloy nanoclusters[19]. Extensive reviews have provided effective evidence for bulk–shell segregation in NaK nanoalloy cluster structures[20, 21].

Atomic composition is affected by binding energy induced changes in geometric structure, energy density, charge transfer, atomic orbital hybridization, and bond strain in the alloy nanocluster[22]. It is desirable to understand which of the possible chemical ordering patterns—bulk–shell, onion layer, or random blend—is the most probable for a particular nanoalloy cluster. To understand chemical ordering or atomic coordination in nanoalloys[23, 24], it can be helpful to analyze a series of compounds with similar geometric structures[25]. To this end, it is necessary to determine the trends of structural variations with chemical ordering.

In this report, we present compositional effects in 55-atom NaK nanoalloys based on the framework of BOLS notation[26] and DFT calculations. This approach allows us not only to reproduce the trend of composition with binding energy, but also clarifies the physical origins of the energy shifts associated with composition in nanoalloys. Further, it allows clarification of the atomic bonding and quantification of the bond energy ratio, coordination number, bond energy density, and local bond strain values in these clusters.

## 2. Principles and methods of calculation

### 2.1 DFT Calculations

In this study, we used DFT to calculate the structures of $Na_xK_{55-x}$ alloy nanoclusters in which $x$, the number of Na atoms in the cluster, takes values from 0 to 55. We then selected several stable representative clusters from these screening results: the $Na_{55}$, $K_{55}$, $Na_7K_{48}$, and $Na_{15}K_{40}$ nanoclusters (**Fig. 1**). These selected preliminary NaK nanoclusters were modeled using the Gupta potentials devised by Andrés Aguado, and optimized to obtain the structures with the lowest energy (global minimum energy geometries)[21]. The first-principles calculations were performed using the Vienna Ab initio Simulation Package code within the projector augmented wave method[27, 28], employing Perdew–Burke–Ernzerhof pseudopotentials[29]. The energy cut-off for the plane-wave was set to 400 eV. Both force convergence thresholds (0.01 eV/Å) and an energy convergence criteria ($10^{-5}$ eV) were used to obtain the optimized geometries. The Brillouin zone was sampled at the Gamma-point only. We used cells with 15-Å vacuum spaces along the $x$, $y$, and $z$ directions for all the calculated structures. The energy without entropy result from each DFT calculation and the corresponding $Na_xK_{55-x}$ cluster symmetry are given in **Table 1**.

### 2.2 BOLS notation concept

According to BOLS theory, atomic undercoordination shortens and strengthens the remaining bonds between the undercoordinated atoms. Chemical bond shrinkage leads to localized atom charges and increases in energy density; the increase in bond energy is associated with the deepening of the local potential well and the shifting of energy levels. The following expressions summarize bond relaxation theory[26]:

$$\begin{cases} d_z = C_i d_b & \text{(bond length)} \\ E_z = C_i^{-m} E_b & \text{(bond energy)} \\ C_i = 2 \big/ \{1 + \exp[(12 - z_i)/(8z_i)]\} & \text{(bond contraction ratio)} \end{cases},$$

$$\tag{1}$$

where $C_i$ represents the bond contraction ratio, $d_z$ is the bond length after atomic relaxation, and $d_b$ is the bond length for the atom in the bulk; $E_b$ represents the energy of a single bond of the atom in the bulk, and $E_z$ is the bond energy of a single bond for the surface atom. The coordination number of the surface atom is $z$ and that of the bulk atom is $z_b$; $z_b = 12$. The $m$ is the bond nature index, and for metals, $m = 1$.

## 3. Results and discussion

### 3.1 Structural and BE shifts

**Fig. 1** shows the global minimum structures of the $Na_xK_{55-x}$ alloy nanoclusters; in the figure, purple spheres indicate Na atoms and light blue spheres are K atoms. Structural parameter information, point group symmetries, binding energies and the energy without entropy values of $Na_xK_{55-x}$ clusters are given in **Table 1**. In the K-rich clusters, i.e. $x = 1–26$, the Na atoms occupy positions within the bulk atomic layers. For $x = 27–42$, The Na atoms move from the bulk positions to positions in the shell atomic layer. For the Na-rich structures, $x = 42–55$, the K atoms occupy positions in the shell atomic layer, and there is no segregation of the K component to the inside of the cluster in this dilute-Na limit.

**Fig. 2** shows the DFT-calculated binding energy values for $Na_xK_{55-x}$ clusters. These results reveal the changes in binding energy due to compositional effects. **Fig. 2a** shows that the $Na_4K_{51}$ cluster has the largest Na 2p binding energy (–25.228 eV) and the $Na_{15}K_{40}$ cluster has the smallest binding energy (–24.753 eV). For $x = 0–4$, the Na 2p binding energy shifts to more negative values with $x$ in the $Na_xK_{55-x}$ alloy nanoclusters. For $x = 4–15$, the binding energy increases with $x$, shifting to more positive values with $x$. For $x = 15–55$, the overall trend in these nanoclusters is for the binding energy to shift negatively with $x$. These results are consistent with the trend of electron affinities and excess energy observed by Aguado et al[21]. **Fig. 2b** reveals that the $Na_{39}K_{16}$ cluster has the largest K 3p binding energy (–15.803 eV), and the $Na_{13}K_{42}$ cluster has the smallest binding energy (–15.523 eV). For $x = 0–4$, the binding energy shifts negatively with $x$, whereas for $x = 4–13$, the overall trend is for the binding energy to increase with $x$. For $x = 13–54$, the trend becomes positive once more, as binding energy shifts to more negative values with $x$. These results are consistent with the trends observed for the Na $2p$ binding energy values.

**3.2 Electronic properties**

To analyze the effects of composition and structure on binding energy, we investigated the electronic properties of the clusters. We selected the $Na_{55}$, $K_{55}$, $Na_7K_{48}$, and $Na_{15}K_{40}$ clusters and determined the density of states (DOS) to understand their electronic structure. **Fig. 3** shows the Na $2p$ and $3p$ binding energies for the $Na_{55}$, $K_{55}$, $Na_7K_{48}$, and $Na_{15}K_{40}$ clusters. In **Fig. 3a**, it is apparent that the Na $2p$ binding energy increases according to the trend: $Na_{55}$ (–25.034 eV) > $Na_7K_{48}$ (–24.99 eV) > $Na_{15}K_{40}$ (–24.753 eV). From **Fig. 3b**, the K $3p$ binding energy trend may be extracted: $K_{55}$ (–15.532 eV) < $Na_7K_{48}$ (–15.592 eV) < $Na_{15}K_{40}$ (–15.603 eV). These results are consistent with the XPS data reported by Tchaplyguine et al. Our DFT-screening approach, therefore, can achieve energy results that agree with measurements of experimentally prepared alloy nanomaterials.

**Fig. 4** presents shell-resolved local density of states (LDOS) plots for spatial components of the $Na_{55}$, $K_{55}$, $Na_7K_{48}$, and $Na_{15}K_{40}$ alloy nanoclusters. In **Fig. 4a**, three components of the $Na_{55}$ cluster, the bulk B and surface $S_2$ and $S_1$ components, with binding energies peaking at –24.839 eV, –25.034 eV, and –25.217 eV, respectively, are resolved. **Fig. 4b** displays the $K_{55}$ LDOS, with peak binding energies of –15.437 eV, –15.570 eV, and –15.712 eV, resolved. **Fig. 4a and 4b** allow us to observe the ordering of the binding energy for the surface and bulk atoms: the binding energies of the surface atoms are larger than those of the bulk atoms. **Fig. 4c** shows the DOS of the Na $2p$ level in the $Na_7K_{48}$ cluster. This cluster has bulk and shell components. The bulk–shell-resolved DOS components have peaks at 24.882 (bulk) and 25.004 (shell) eV. **Fig. 4d** shows the DOS of the Na $2p$ level of the $Na_{15}K_{40}$ cluster. The resolved components peak at 24.748 (bulk) and 24.87 (shell) eV. These results demonstrate that the binding energies of the atoms in the shell are larger than those of the bulk atoms.

**Fig. 4e** displays the DOS for the K $3p$ level of the $Na_7K_{48}$ cluster. This cluster has bulk, B, surface, $S_2$ and $S_1$, components, resolved with peaks at 15.354, 15.580, and 15.776 eV, respectively. **Fig. 4f** shows the DOS of the K 3*p* level of the Na$_{15}$K$_{40}$ nanocluster; the bulk–shell-resolved components have peaks at 15.232, 15.535, and 15.789 eV. These decompositions presented here include the principal structural components of each cluster, but the complex cluster geometries obtained could perhaps be decomposed into a larger number of components with different binding energies for the different elements.

**3.3 Bond properties**

According to the BOLS correlation, different compositions and structures have different binding energies. For a given DFT-calculated binding energy shift, the relationship between the energy level *v* of the energy shift $\Delta E_v(i)$ and the atomic coordination number *z* is as follows[30]:

$$\begin{cases} z = \dfrac{12}{\left\{8\ln\left(\dfrac{2\Delta E_v'(i) + \Delta E_v(12)}{\Delta E_v(12)}\right) + 1\right\}} \\ E_v(i) = \Delta E_v'(i) + E_v(12) \end{cases}$$

(2)

In Eq. (2), $\Delta E_v'(i) = E_v(i) - E_v(12)$ represents the binding energy shift for an atom in the ideal bulk. Quantitative information on the bulk shifts $\Delta E_v(12)$ of Na (2.401 eV) and K (2.754 eV) are available from the Na(110) and K(110) surfaces via XPS measurements[30].

Using the bond order (*z*), length (*d*), energy (*E*), and nature (*m*) parameters, we can predict the bond energy ratio $\gamma$, bond energy density $\delta E_d$, and local bond strain $\varepsilon_z$ as follows[31]:

$$\begin{cases} \gamma = E_z / E_b = C_i^{-m} = (\Delta E_v(12) + \Delta E_v'(i))/\Delta E_v(12) & \text{(bond energy ratio)} \\ \delta E_d(z) = (E_z / d_i^3)/(E_b / d_b^3) = C_i^{-m-3} = \gamma^4 & \text{(bonding energy density)} \\ \varepsilon_z = d_z / d_b - 1 = \gamma^{-1} - 1 & \text{(local bond strain)} \end{cases}$$

(3)

The bond energy ratio parameter, $\gamma$, represents the change in bond energy or strength.

If $\gamma < 1$, the bond energy $E_z$ is reduced, i.e. the bond is weakened, and if $\gamma > 1$, the bond energy increases and the bond becomes stronger. Bond energy density, $\delta E_d$, represents the amount of electron localization. Local bond strain, $\varepsilon_z$, represents the contraction ratio of the atomic bond length $d_z$.

**Eq. (2)** can be used to calculate the coordination number, $z$, of an atom from its binding energy. The results of such calculations are shown in **Table 2** for selected $Na_xK_{55-x}$ alloy nanoclusters. Calculations carried out using **Eq. (3)** reveal that atoms with lower coordination numbers have higher binding energies and bond energy densities. An increase in bond energy causes an increase in local bond strain and bond energy density, as illustrated in **Fig. 5**. These results indicate that the composition and structure of an $Na_xK_{55-x}$ alloy nanocluster determines its binding energy shift.

## 4. Conclusion

We use first principles calculations to understand compositional effects in $Na_xK_{55-x}$ alloy clusters. In connection with BOLS theory, we found that compositional effects in $Na_xK_{55-x}$ nanoclusters are principally determined by the arrangement of the different atoms within its structure. Atoms at different positions in the cluster have different binding energies. Further, the location of the atoms determines the physical and chemical properties of the material. We described the quantitative relationships linking atomic coordination number $z$, bond energy density $E_d$, binding energy shifts $\Delta E'_v(z)$, and local bond strain $-\varepsilon_i$ for the alloy. These results should provide a theoretical reference for the experimental preparation of high-performance alloy nanoclusters.

**Figures and tables with captions**

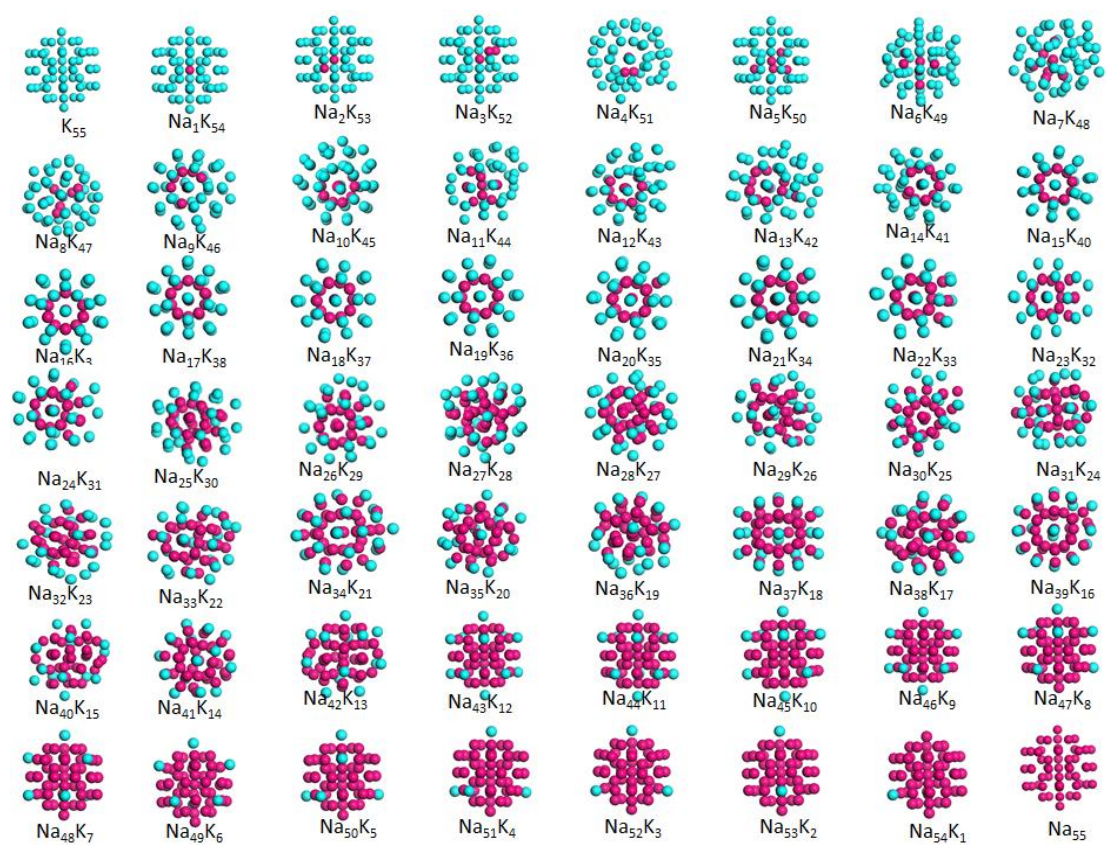

**Fig. 1** Optimized structures of Na$_x$K$_{55-x}$ clusters determined using the initial configurations of Andrés Aguado and Gupta potentials[19]. Purple and light blue spheres represent Na and K atoms, respectively. Table 1 summarizes the derived numeric information for these optimized structures.

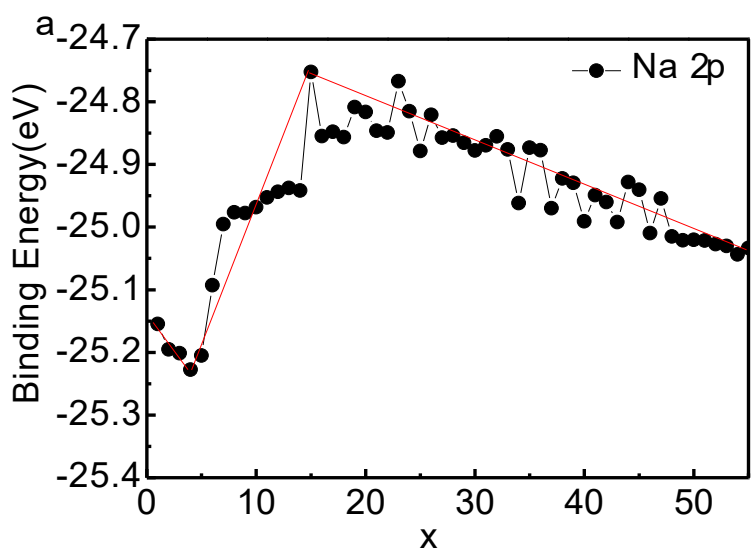

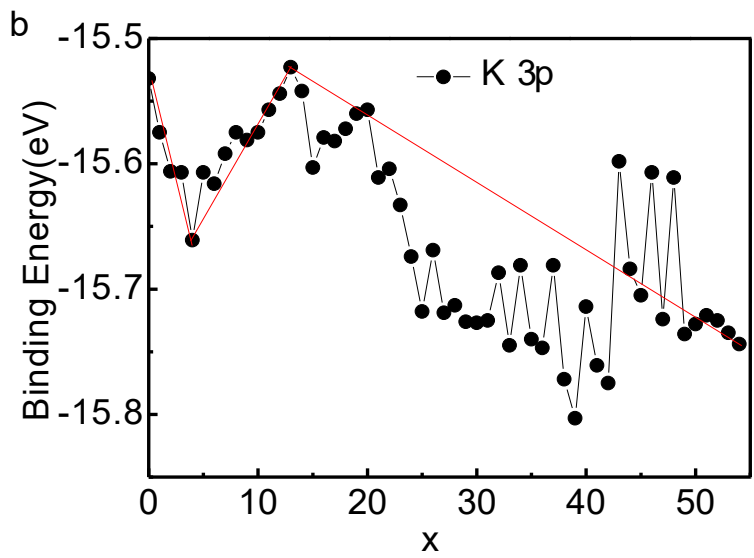

**Fig. 2** Binding energies of Na$_x$K$_{55-x}$ clusters for selected compositions.

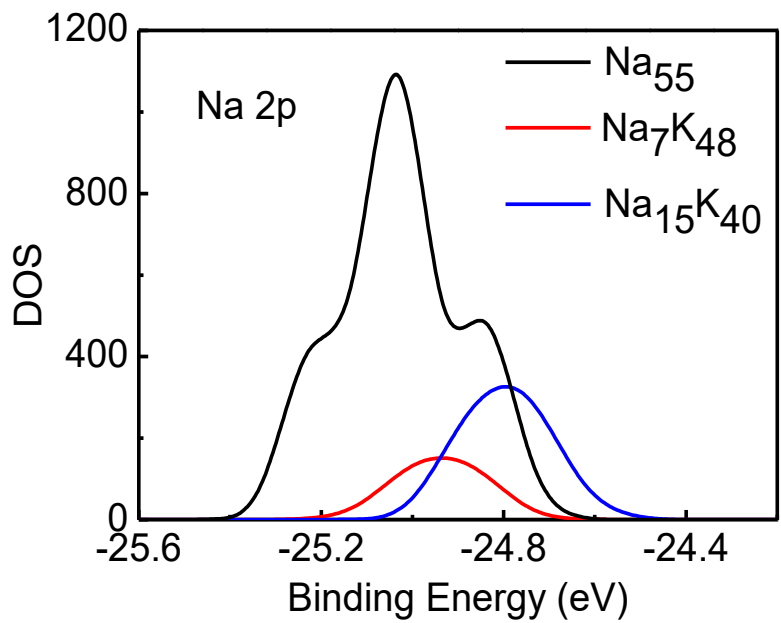

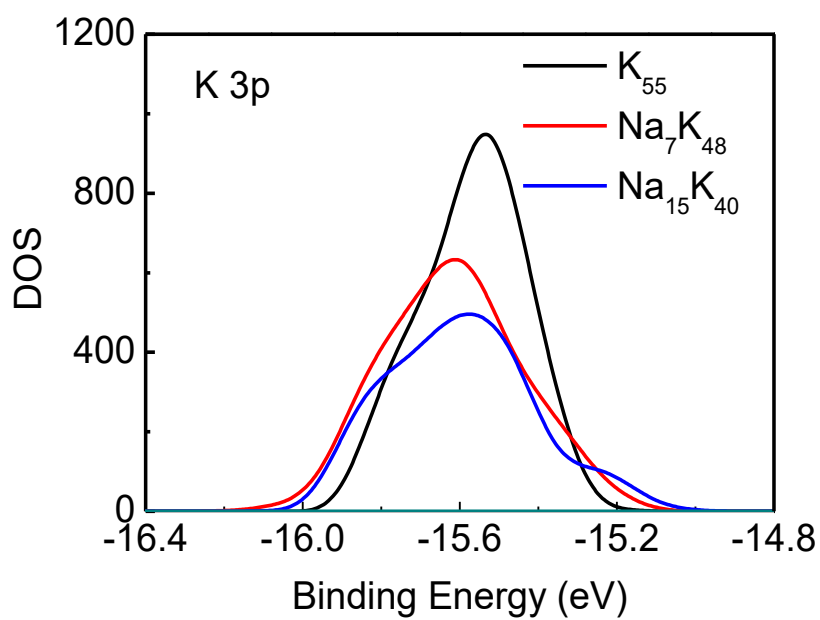

**Fig. 3** Binding energies of Na 2p (top) and K 3p (bottom) levels in $Na_{55}$, $K_{55}$, $Na_7K_{48}$, and $Na_{15}K_{40}$ clusters.

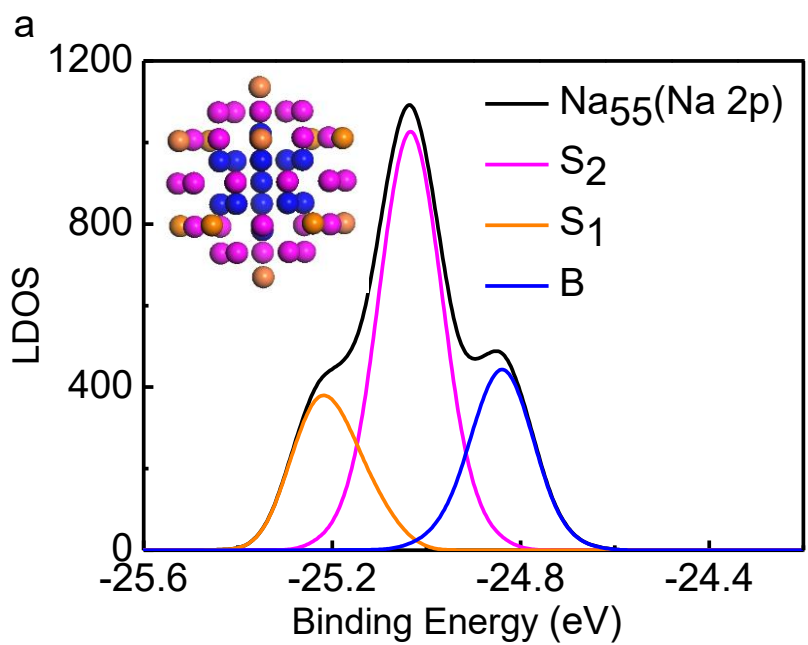

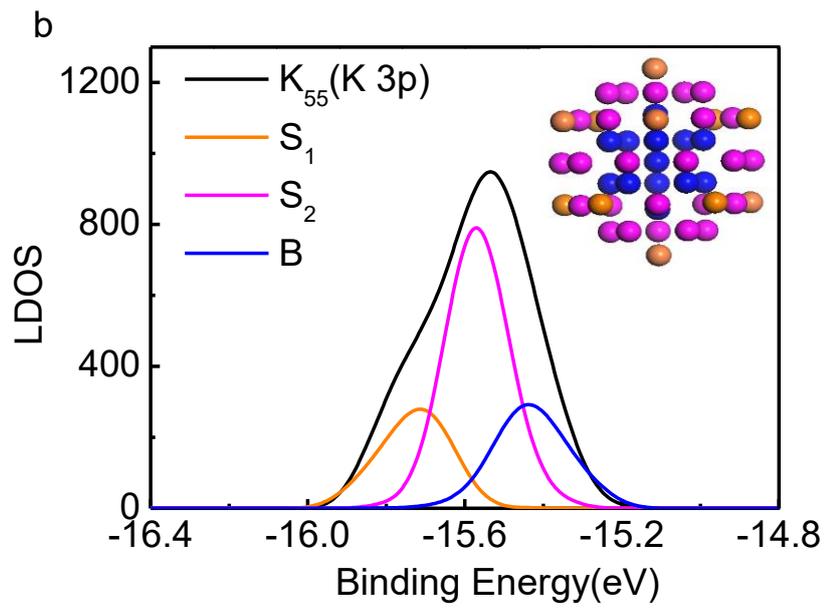

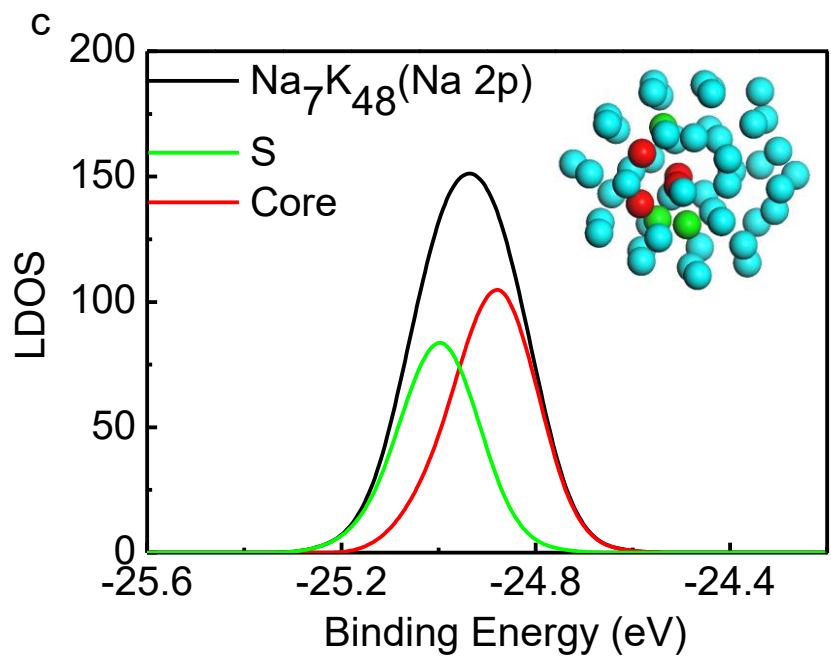

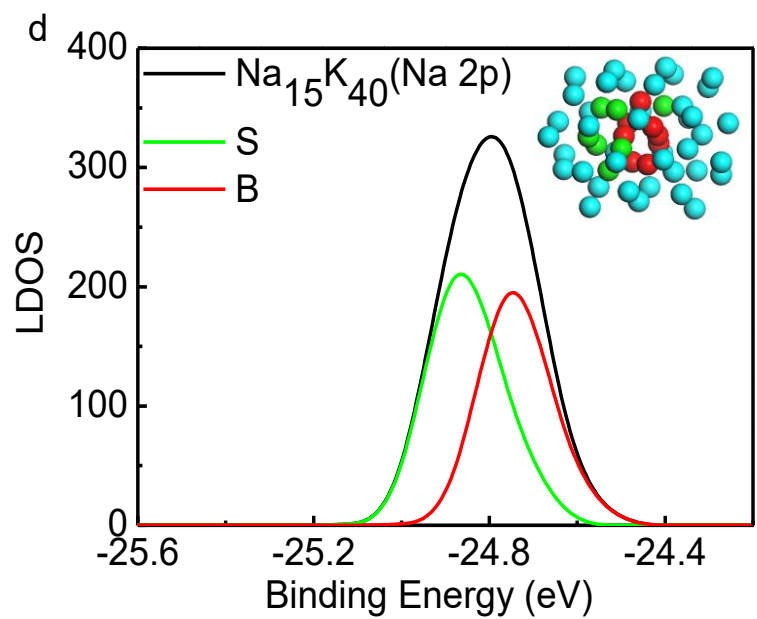
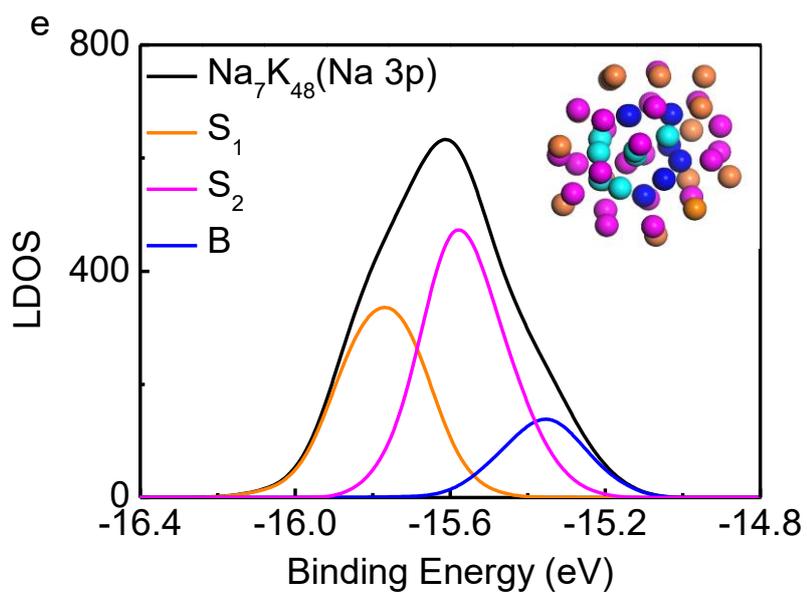

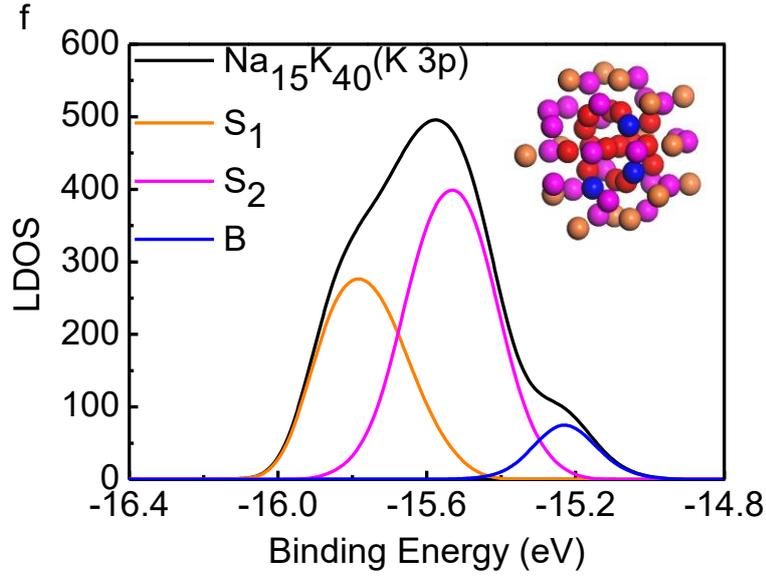

**Fig. 4** Bulk–shell-resolved local density of states plots for (a) $Na_{55}$ (Na 2$p$), (b) $K_{55}$ (K 3$p$), (c) $Na_7K_{48}$ (Na 2$p$), (d) $Na_{15}K_{40}$ (Na 2$p$), (e) $Na_7K_{48}$ (K 3$p$), and (f) $Na_{15}K_{40}$ (K 3$p$) clusters. S–shell; B–bulk; $S_1$–shell component 1; $S_2$–shell component 2.

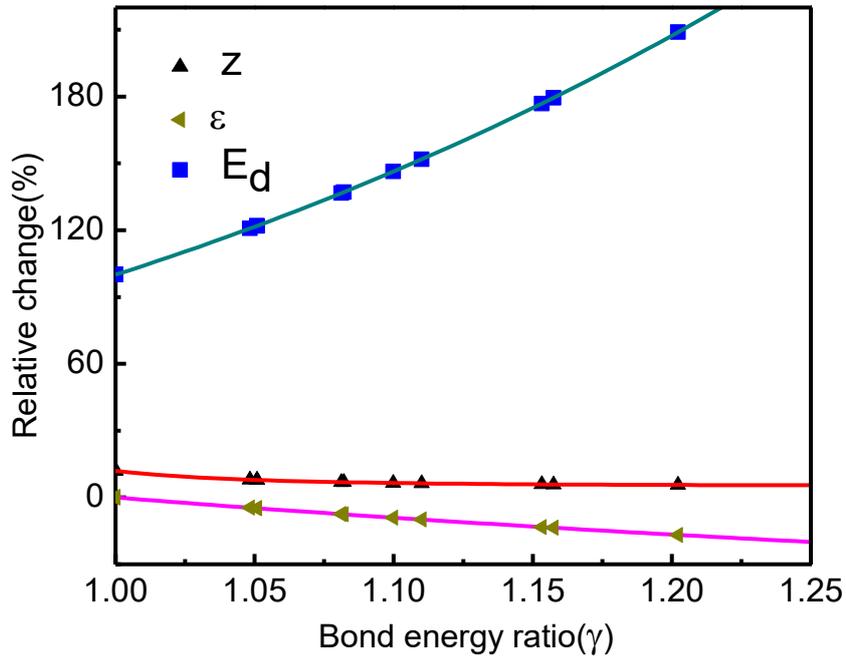

**Fig. 5** Coordination number ($z$), local bond strain ($-\varepsilon_z$), and bond energy density ($E_d$) for selected $Na_xK_{55-x}$ clusters. Table 2 shows the derived information.

**Table 1** Symmetry group, binding energy, and energy without entropy of $Na_xK_{55-x}$ clusters.

| | Symmetry group | Energy (eV) | Na (2p) binding energy (eV) | K (3p) binding energy (eV) |
|---|---|---|---|---|
| $Na_1K_{54}$ | $I_h$ | -53.595 | -25.155 | -15.575 |
| $Na_2K_{53}$ | $C_{5v}$ | -53.802 | -25.195 | -15.606 |
| $Na_3K_{52}$ | $C_{2v}$ | -54.006 | -25.202 | -15.607 |
| $Na_4K_{51}$ | $C_s$ | -54.203 | -25.228 | -15.661 |
| $Na_5K_{50}$ | $C_s$ | -54.399 | -25.205 | -15.607 |
| $Na_6K_{49}$ | $C_s$ | -54.674 | -25.093 | -15.616 |
| $Na_7K_{48}$ | $C_1$ | -55.357 | -24.995 | -15.592 |
| $Na_8K_{47}$ | $C_s$ | -55.729 | -24.976 | -15.575 |
| $Na_9K_{46}$ | $C_1$ | -56.189 | -24.978 | -15.581 |
| $Na_{10}K_{45}$ | $C_1$ | -56.571 | -24.968 | -15.575 |
| $Na_{11}K_{44}$ | $C_1$ | -56.969 | -24.953 | -15.557 |
| $Na_{12}K_{43}$ | $C_1$ | -57.392 | -24.944 | -15.544 |
| $Na_{13}K_{42}$ | $C_1$ | -57.764 | -24.938 | -15.523 |
| $Na_{14}K_{41}$ | $C_1$ | -58.121 | -24.942 | -15.542 |
| $Na_{15}K_{40}$ | $C_s$ | -58.269 | -24.753 | -15.603 |
| $Na_{16}K_{39}$ | $C_s$ | -58.676 | -24.855 | -15.579 |
| $Na_{17}K_{38}$ | $C_s$ | -59.085 | -24.848 | -15.582 |
| $Na_{18}K_{37}$ | $C_{5v}$ | -60.593 | -24.857 | -15.572 |
| $Na_{19}K_{36}$ | $C_{5v}$ | -59.804 | -24.809 | -15.560 |
| $Na_{20}K_{35}$ | $C_s$ | -60.037 | -24.817 | -15.557 |
| $Na_{21}K_{24}$ | $C_s$ | -60.270 | -24.846 | -15.611 |
| $Na_{22}K_{33}$ | $C_s$ | -60.494 | -24.849 | -15.604 |
| $Na_{23}K_{32}$ | $C_1$ | -60.688 | -24.767 | -15.633 |
| $Na_{24}K_{31}$ | $C_1$ | -60.811 | -24.815 | -15.674 |
| $Na_{25}K_{30}$ | $C_1$ | -60.651 | -24.878 | -15.718 |
| $Na_{26}K_{29}$ | $C_s$ | -60.816 | -24.821 | -15.669 |
| $Na_{27}K_{28}$ | $C_1$ | -61.065 | -24.857 | -15.719 |
| $Na_{28}K_{27}$ | $C_s$ | -61.333 | -24.854 | -15.713 |
| $Na_{29}K_{26}$ | $C_s$ | -61.548 | -24.866 | -15.726 |
| $Na_{30}K_{25}$ | $C_1$ | -61.787 | -24.878 | -15.727 |
| Na$_{31}$K$_{24}$ | C$_1$ | -61.990 | -24.870 | -15.725 |
| Na$_{32}$K$_{23}$ | C$_1$ | -62.199 | -24.855 | -15.687 |
| Na$_{33}$K$_{22}$ | C$_1$ | -62.474 | -24.876 | -15.745 |
| Na$_{34}$K$_{21}$ | C$_{2v}$ | -62.785 | -24.962 | -15.681 |
| Na$_{35}$K$_{20}$ | C$_1$ | -63.029 | -24.873 | -15.740 |
| Na$_{36}$K$_{19}$ | C$_1$ | -63.263 | 24.877 | -15.747 |
| Na$_{37}$K$_{18}$ | Cs | -63.522 | -24.970 | -15.681 |
| Na$_{38}$K$_{17}$ | C$_{2v}$ | -63.912 | -24.923 | -15.772 |
| Na$_{39}$K$_{16}$ | C$_1$ | -63.856 | -24.930 | -15.803 |
| Na$_{40}$K$_{15}$ | C$_1$ | -64.205 | -24.991 | -15.714 |
| Na$_{41}$K$_{14}$ | C$_1$ | -64.399 | -24.949 | -15.761 |
| Na$_{42}$K$_{13}$ | Cs | -64.663 | -24.960 | -15.775 |
| Na$_{43}$K$_{12}$ | I$_h$ | -64.573 | -24.992 | -15.598 |
| Na$_{44}$K$_{11}$ | C$_{5v}$ | -64.838 | -24.928 | -15.684 |
| Na$_{45}$K$_{10}$ | C$_{2v}$ | -65.099 | -24.941 | -15.705 |
| Na$_{46}$K$_9$ | C$_{3v}$ | -65.352 | -25.006 | -15.607 |
| Na$_{47}$K$_8$ | Cs | -65.608 | -24.954 | -15.724 |
| Na$_{48}$K$_7$ | Cs | -65.862 | -25.015 | -15.611 |
| Na$_{49}$K$_6$ | C$_{5v}$ | -66.14 | -25.021 | -15.736 |
| Na$_{50}$K$_5$ | Cs | -66.359 | -25.020 | -15.728 |
| Na$_{51}$K$_4$ | C$_{2v}$ | -66.602 | -25.022 | -15.721 |
| Na$_{52}$K$_3$ | C$_{3v}$ | -66.841 | -25.027 | -15.725 |
| Na$_{53}$K$_2$ | C$_{2v}$ | -67.046 | -25.030 | -15.735 |
| Na$_{54}$K$_1$ | C$_{5v}$ | -67.311 | -25.044 | -15.744 |
| Na$_{55}$ | I$_h$ | -67.542 | -25.034 | --- |
| K$_{55}$ | I$_h$ | -53.209 | --- | -15.532 |

**Table 2** BOLS-derived binding energy shifts $\Delta E'_v(z) = \Delta E_v(z) - \Delta E_v(B)$, coordination number ($z$), bond energy ratio ($\gamma$; %), bond energy density ($E_d$; %), and local bond strain ($-\varepsilon_z$; %) for selected $Na_xK_{55-x}$ clusters.

|  | $i$ | $E_v(z)$ (eV) | $\Delta E'_v(z)$ | $z$ | $\gamma$ | $-\varepsilon_z$ | $\delta E_d$ |
|---|---|---|---|---|---|---|---|
| $Na_{55}$(Na) | $S_1$ | –25.217 | 0.378 | 3.76 | 1.157 | 13.60 | 179.47 |
| (Na) | $S_2$ | –25.034 | 0.195 | 5.44 | 1.081 | 7.51 | 136.66 |
| (Na) | B | –24.839 | 0 | 12.00 | 1 | 0.00 | 100 |
| $K_{55}$(K) | $S_1$ | –15.712 | 0.275 | 4.89 | 1.100 | 9.08 | 146.33 |
| (K) | $S_2$ | –15.570 | 0.133 | 6.91 | 1.048 | 4.61 | 120.76 |
| (K) | B | –15.437 | 0 | 12.00 | 1 | 0.00 | 100 |
| $Na_7K_{48}$(K) | $S_1$ | –15.776 | 0.422 | 3.82 | 1.153 | 13.29 | 176.87 |
| (K) | $S_2$ | –15.580 | 0.266 | 5.42 | 1.082 | 7.58 | 137.09 |
| (K) | B | –15.354 | 0 | 12.00 | 1 | 0.00 | 100.00 |
| $Na_7K_{48}$ (Na) | $S_1$ | –25.004 | 0.122 | 6.76 | 1.051 | 4.84 | 121.93 |
| (Na) | B | –24.882 | 0 | 12.00 | 1 | 0.00 | 100 |
| $Na_{15}K_{40}$(K) | $S_1$ | –15.789 | 0.557 | 3.23 | 1.202 | 16.82 | 108.92 |
| (K) | $S_2$ | –15.535 | 0.303 | 4.63 | 1.110 | 9.91 | 151.82 |
| (K) | B | –15.232 | 0 | 12.00 | 1 | 0.00 | 100 |
| $Na_{15}K_4$ (Na) | $S_1$ | –24.870 | 0. 122 | 6.76 | 1.051 | 4.84 | 121.93 |
| (Na) | B | –24.748 | 0 | 12.00 | 1 | 0.00 | 100 |